\documentclass[pre,twocolumn,amsmath,amssymb,floatfix,superscriptaddress,showpacs]{revtex4}

\usepackage{graphicx}
\usepackage{latexsym}
\usepackage{amsmath}
\usepackage{amssymb}
\usepackage{amsfonts}
\usepackage{bm}

\newcommand{\la}{\left<}
\newcommand{\ra}{\right>}

\newcommand{\rvec}{\ensuremath{\underline{r}}}

\newcommand{\ddiff}{\ensuremath{\text{d}}}

\newcommand{\kB}{\mbox{$k_{\rm B}$}}

\newcommand{\NVgT}{\ensuremath{\text{NV}\gamma\text{T}}}
\newcommand{\NVtT}{\ensuremath{\text{NV}\tau\text{T}}}
\newcommand{\MA}{\ensuremath{M_\mathrm{A}}}
\newcommand{\muA}{\ensuremath{\mu_\mathrm{A}}}
\newcommand{\muAhat}{\ensuremath{\hat{\mu}_\mathrm{A}}}
\newcommand{\muF}{\ensuremath{\sigma}}
\newcommand{\muFtild}{\ensuremath{\tilde{\sigma}}}
\newcommand{\muFstar}{\ensuremath{\sigma_{\star}}}

\newcommand{\muFtau}{\ensuremath{\sigma|_{\tau}}}
\newcommand{\Cttild}{\ensuremath{\tilde{c}(t)}}
\newcommand{\Ctild}{\ensuremath{\tilde{c}}}

\newcommand{\Geq}{G_\mathrm{eq}}
\newcommand{\GF}{g}
\newcommand{\GFtild}{\tilde{g}}

\newcommand{\tauhat}{\hat{\tau}}
\newcommand{\gamhat}{\hat{\gamma}}

\newcommand{\Hexhat}{\hat{\cal H}_\mathrm{ex}}

\newcommand{\dtMD}{\delta t_\mathrm{MD}}
\newcommand{\tauA}{t_\mathrm{A}}

\newcommand{\tstar}{t_{\star}}

\newcommand{\tsamp}{\Delta t}
\newcommand{\ttraj}{t_\mathrm{traj}}

\begin{document}

\title{Simple-average expressions for shear-stress relaxation modulus}

\author{J.P.~Wittmer}
\email{joachim.wittmer@ics-cnrs.unistra.fr}
\affiliation{Institut Charles Sadron, Universit\'e de Strasbourg \& CNRS, 23 rue du Loess, 67034 Strasbourg Cedex, France}
\author{H.~Xu}
\affiliation{LCP-A2MC, Institut Jean Barriol, Universit\'e de Lorraine \& CNRS, 1 bd Arago, 57078 Metz Cedex 03, France}
\author{J. Baschnagel}
\affiliation{Institut Charles Sadron, Universit\'e de Strasbourg \& CNRS, 23 rue du Loess, 67034 Strasbourg Cedex, France}

\begin{abstract}
Focusing on isotropic elastic networks we propose a novel simple-average expression $G(t) = \muA - h(t)$ 
for the computational determination of the shear-stress relaxation modulus $G(t)$
of a classical elastic solid or fluid and its equilibrium modulus $\Geq = \lim_{t \to \infty} G(t)$. 
Here, $\muA = G(0)$ characterizes the shear transformation of the system at $t=0$
and $h(t)$ the (rescaled) mean-square displacement of the instantaneous shear stress $\tauhat(t)$
as a function of time $t$.
While investigating sampling time effects we also discuss the related expressions in terms 
of shear-stress autocorrelation functions.
We argue finally that our key relation may be readily adapted for more general linear response functions. 
\end{abstract}
\pacs{05.70.-a,05.20.Gg,83.10.Ff,62.20.D-,83.80.Ab}
\date{\today}
\maketitle

\begin{figure}[t]
\centerline{\resizebox{0.9\columnwidth}{!}{\includegraphics*{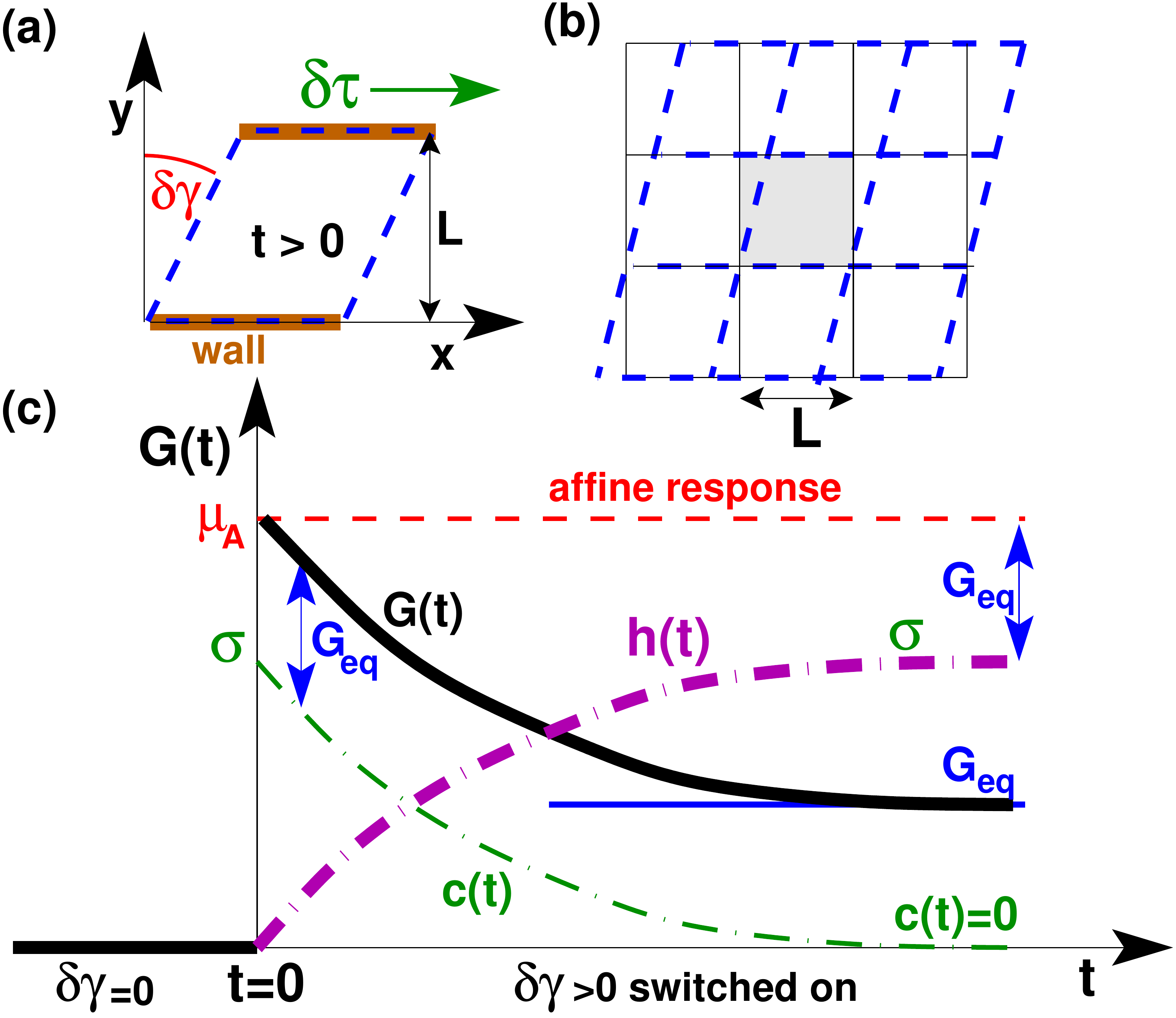}}}
\caption{Notations and addressed problem:
{\bf (a)} experimental setup with $\delta \gamma$ being the shear strain,
$\delta \tau$ the shear stress increment and $L$ the system size,
{\bf (b)} affine plain shear with periodic boundary conditions,
{\bf (c)} sketch of Eq.~(\ref{eq_keyPRE}) and Eq.~(\ref{eq_key})
with $G(t)$ being the response modulus (bold solid line), 
$\Geq$ the equilibrium shear modulus (thin solid horizontal line),
$\muA$ the affine shear-elasticity (thin dashed horizontal line), 
$\muF$ the shear-stress fluctuations, 
$c(t)$ the  ACF (thin dash-dotted line) and $h(t)=\muF-c(t)$
the MSD (bold dash-dotted line). 
\label{fig_sketch}
}
\end{figure}

\section{Introduction}
\label{sec_intro}

\paragraph*{Background.}
A central rheological property characterizing the linear response of (visco)elastic bodies
is the shear relaxation modulus $G(t)$ \cite{RubinsteinBook,WittenPincusBook,DoiEdwardsBook}.
Assuming for simplicity an isotropic body, $G(t) = \delta \tau(t)/\delta \gamma$
may be obtained from the stress increment $\delta \tau(t)$ after a small step strain $\delta \gamma$ 
has been imposed at time $t=0$ as sketched in panel (a) of Fig.~\ref{fig_sketch}.
The instantaneous shear stress $\tauhat(t)$ may be determined 
in a numerical study, as shown in panel (b) for a sheared periodic simulation box,
from the model Hamiltonian and the particle positions and momenta 
\cite{AllenTildesleyBook}.
The long-time response yields the equilibrium shear modulus 
$\Geq = \lim_{t\to \infty} G(t)$ as shown in panel (c).
More readily, one may obtain $\Geq$ by means of equilibrium simulations performed at constant volume $V$ and 
shear strain $\gamma$ using the stress-fluctuation relation 
\cite{Hoover69,Barrat88,Barrat06,Lutsko89,Szamel15,WXP13,WXB15,WKB15}
\begin{equation}
\Geq = \GF \equiv \muA - \muF 
\label{eq_SFF}
\end{equation}
with $\muA$ being the ``affine shear-elasticity" \cite{WXP13,WXB15,WKB15},
a simple average characterizing the second order energy change under a canonical-affine shear strain \cite{WKB15}.
The second contribution $\muF \equiv \beta V \langle \delta \tauhat^2 \rangle = \muFtild - \muFstar$
stands for the rescaled shear stress fluctuation with $\beta$ being the inverse temperature and 
where we have introduced for later convenience the two terms $\muFtild \equiv \beta V \langle \tauhat^2 \rangle$
and $\muFstar \equiv \beta V \langle \tauhat \rangle^2$.
As shown in Refs.~\cite{WXP13,WXB15,WKB15} Eq.~(\ref{eq_SFF}) can be derived
using the general transformation relation for fluctuations between conjugated ensembles \cite{Lebowitz67}.
Using these transforms for the shear-stress autocorrelation function (ACF)
$c(t) \equiv \beta V \la \delta \tauhat(t) \delta \tauhat(0) \ra \equiv \Cttild - \muFstar$,
Eq.~(\ref{eq_SFF}) can be extended into the time domain \cite{WXB15,WKB15}.
This allows the determination of the shear relaxation modulus using 
\begin{equation}
G(t) = \Geq + c(t) 
\label{eq_keyPRE}
\end{equation}
as illustrated by the thin dash-dotted line in panel (c).
Note that Eq.~(\ref{eq_keyPRE}) is more general than the relation $G(t) = \Cttild$ 
commonly used for liquids \cite{DoiEdwardsBook,HansenBook,Klix12,Szamel15}.
One important consequence of Eq.~(\ref{eq_keyPRE}) is that a {\em finite} shear modulus $\Geq$ is only 
probed by $G(t)$ on time scales where $c(t)$ actually vanishes. While $\Geq$ can be obtained from 
Eq.~(\ref{eq_SFF}), this is not possible using {\em only} $c(t)$ or $\Cttild$ \cite{WXB15}.

\paragraph*{Key points made.}
Note that both Eq.~(\ref{eq_SFF}) and Eq.~(\ref{eq_keyPRE}) assume that the sampling time
$\tsamp$ is much larger than the longest, terminal relaxation time $\tstar$ of the system
and that, hence, time and ensemble averages are equivalent. Since both relations are formulated 
in terms of fluctuating properties, not in terms of ``simple averages" \cite{AllenTildesleyBook},
this suggests that they might converge slowly with increasing $\tsamp$ to their respective
thermodynamic limits. The aim of the present study is to rewrite and generalize (where necessary) 
both relations in terms of simple averages allowing an accurate determination even for $\tsamp \ll \tstar$.
As we shall demonstrate this can be achieved by rewriting Eq.~(\ref{eq_keyPRE}) simply as 
\begin{equation}
G(t) = \muA-\muFtild + \Cttild = G(0) - h(t)
\label{eq_key}
\end{equation}
with $G(0) = \muA$ and $h(t) = \beta V/2 \ \langle (\tauhat(t)-\tauhat(0))^2 \rangle$
being the shear-stress mean-square displacement (MSD).
We have used in the second step of Eq.~(\ref{eq_key}) the exact identity 
\begin{equation}
h(t) = \Ctild(0) - \Ctild(t) = c(0) - c(t)
\label{eq_htCt}
\end{equation}
with $\Ctild(0)=\muFtild$ and $c(0)=\muF$.
Both expressions given in Eq.~(\ref{eq_key}) are numerically equivalent $\tsamp$-independent 
simple averages. 

\paragraph*{Outline.}
We begin by presenting in Sec.~\ref{sec_algo} the numerical model 
and remind some properties of the specific elastic network investigated
already described elsewhere \cite{WXB15,WKB15}.
Our numerical results are then discussed in Sec.~\ref{sec_simu}.
Carefully stating the subsequent time and ensemble averages performed 
we present in Sec.~\ref{simu_static} the pertinent static properties
as a function of the sampling time $\tsamp$. We emphasize in Sec.~\ref{simu_MSD}
that the MSD $h(t)$ is a simple average not explicitly depending on the sampling time 
and on the thermodynamic ensemble. It will be demonstrated that our key relation 
Eq.~(\ref{eq_key}) is a direct consequence of this simple-average behavior. 
We shall then turn in Sec.~\ref{simu_ACF} to the scaling of the ACF $c(t)$.
We come back to the $\tsamp$-dependence of some static properties in Sec.~\ref{simu_compare} where we compare 
several methods for the computation of $\Geq$.  
Our work is summarized in Sec.~\ref{sec_conc} where we discuss some
consequences for the liquid limit and outline finally 
how Eq.~(\ref{eq_key}) may be adapted for more general response functions. 

\section{Some algorithmic details}
\label{sec_algo}

\paragraph*{Model Hamiltonian.}
As in previous work \cite{WXP13,WXB15,WKB15} we illustrate the suggested general relations 
by molecular dynamics (MD) simulations \cite{AllenTildesleyBook} of a periodic 
two-dimensional network of ideal harmonic springs of interaction energy
$\Hexhat = \frac{1}{2} \sum_l K_l \left(r_l - R_l\right)^2$
with $K_l$ being the spring constant, $R_l$ the reference length and $r_l = |\rvec_i - \rvec_j|$
the length of spring $l$. (The sum runs over all springs $l$ between topologically connected 
vertices $i$ and $j$ of the network at positions $\rvec_i$ and $\rvec_j$.)
The mass $m$ of the (monodisperse) particles and Boltzmann's constant $\kB$
are set to unity and Lennard-Jones (LJ) units are assumed throughout this paper.

\paragraph*{Specific network.}
As explained elsewhere \cite{WXP13,WKB15} our network has been constructed using the
dynamical matrix of a strongly polydisperse LJ bead glass quenched down to $T=0$ using 
a constant quenching rate and imposing a relatively large average normal pressure $P=2$. 
This yields systems of number density $\rho \approx 0.96$, 
i.e. linear length $L \approx 102.3$ for the periodic square box.
Since the network topology is by construction {\em permanently fixed}, the shear response $G(t)$
must approach a finite shear modulus $\Geq$ for $t \to \infty$ for all temperatures at 
variance to systems with plastic rearrangements.
If not stated otherwise below, we use an $\NVgT$-ensemble with constant
particle number $N=10^4$, volume $V=L^2$, shear strain $\gamma=0.00071$ \cite{foot_gammazero} 
and temperature $T=1/\beta=0.001$. 
Due to the low temperature the ideal contributions to the average shear stress $\tau$ 
or the affine shear-elasticity $\muA$ are negligible compared to the excess contributions.
The static (ensemble averaged) thermodynamic properties of our finite-temperature network 
relevant for the present study are $P \approx 2$, $\tau \approx \muFstar \approx 0$,
$\muF \approx \muFtild \approx 18$, $\muA \approx 34$ and $\Geq = \muA - \muF \approx 16$
as already shown elsewhere \cite{WXP13,WKB15}.

\paragraph*{Technicalities.}
As in Ref.~\cite{WKB15} the data have been obtained using a Langevin thermostat of friction 
constant $\zeta=1$ and a tiny velocity-Verlet time step $\dtMD = 10^{-4}$.
The instantaneous shear stress $\tauhat$ and several other useful properties
such as the instantaneous affine shear-elasticity $\muAhat$ \cite{WKB15}
are written down every $10 \dtMD$ over several trajectories of length $\ttraj=10^5$.
This is much larger than the longest stress relaxation time $\tstar \approx 10$
of the network properly defined in Sec.~\ref{simu_static}.
Packages of various sampling times $\tsamp \le \ttraj$, as shown in Fig.~\ref{fig_static},
are then analyzed in turn using first time (gliding) averages within each package \cite{AllenTildesleyBook}. 
Finally, we ensemble-average over different $\tsamp$-packages.

\section{Computational results}
\label{sec_simu}

\subsection{Sampling-time dependence of static properties}
\label{simu_static}

\paragraph*{Notations.}
A time average of a property $a$ within a $\tsamp$-package is denoted below by a horizontal bar, 
$\overline{a}$, and an ensemble average by $\langle a \rangle$. While ``simple averages" of form 
$\langle \overline{a} \rangle$ generally do not depend on the sampling time, this may be different for 
averages of type $\langle \overline{a}^2 \rangle$ and similar non-linear ``fluctuations".
To mark this sampling time dependence we often note $\tsamp$ as an (additional) argument for the relevant property.
Obviously, ergodicity implies $\overline{a} \to \langle a \rangle$ for large sampling times
$\tsamp \gg \tstar$ for all properties considered. Hence, $\langle \overline{a}^2 \rangle \to
\langle \langle a \rangle^2 \rangle = \langle a \rangle^2$, i.e. all $\tsamp$-effects must ultimately become irrelevant
and the argument $\tsamp$ is dropped again to emphasize the thermodynamic limit.

\begin{figure}[t]
\centerline{\resizebox{1.0\columnwidth}{!}{\includegraphics*{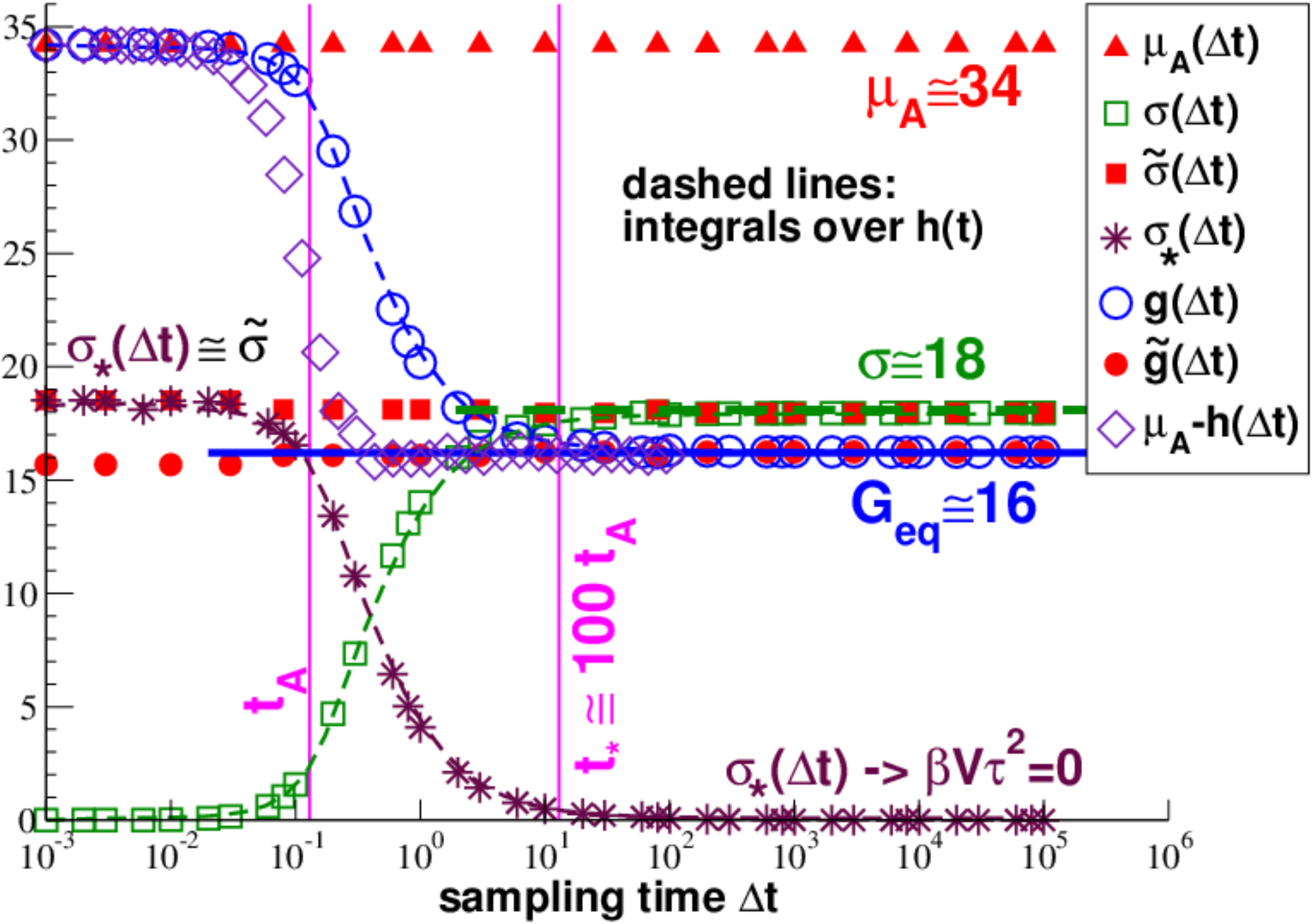}}}
\caption{Sampling time dependence of various static properties.
The ``simple averages" $\muA$, $\muFtild$ and $\GFtild=\muA-\muFtild$ 
do not dependent on $\tsamp$ (filled symbols), while $\muFstar(\tsamp)$, $\muF(\tsamp)$ and $\GF(\tsamp)$ 
do for $\tsamp \ll \tstar \approx 10 \approx 100 \tauA$. Only for $\tsamp \gg \tstar$ these 
``fluctuations" converge to their thermodynamic limits $\muFstar=0$, $\muF\approx 18$ and $\GF = \Geq \approx 16$.
The dashed lines indicate for $\muFstar(\tsamp)$, $\muF(\tsamp)$ and $\GF(\tsamp)$
the respective integrals over the MSD $h(t)$ according to Eq.~(\ref{eq_muFtsamp_ht}).
The diamonds represent $\muA-h(t=\tsamp)$ ensemble-averaged over $10^4$ $\tsamp$-packages.
\label{fig_static}
}
\end{figure}

\paragraph*{Simple averages.}
We begin by verifying the (slightly trivial) $\tsamp$-dependence of the simple averages
\begin{eqnarray}
\muA(\tsamp) & \equiv & \la \overline{\muAhat} \ra \label{eq_muA_def} \\
\muFtild(\tsamp) & \equiv & \beta V \la \overline{\tauhat^2} \ra \label{eq_muFtild_def} \\
\GFtild(\tsamp) & \equiv & \muA(\tsamp) - \muFtild(\tsamp) \label{eq_GFtild_def}
\end{eqnarray}
as shown in Fig.~\ref{fig_static} (filled symbols). As expected, all three simple averages do indeed not depend 
on the sampling time and we shall drop below the argument $\tsamp$. 
We remind that according to the stress-fluctuation formula Eq.~(\ref{eq_SFF}) $\muA$ gives an 
upper bound for the shear-modulus $\Geq$ \cite{WXP13}, while the rescaled second shear stress moment $\muFtild$ gives the 
leading term to the shear-stress fluctuation $\muF(\tsamp)$.
Consistently with other work \cite{Barrat06,WXP13} 
$\muA$ is about twice as large as $\muFtild$ and $\GFtild$ is thus finite.
As can be seen, $\GFtild$ is essentially identical (for reasons given below) to the shear modulus 
$\Geq \approx 16$ indicated by the bold solid line.

\paragraph*{Fluctuations.}
While the simple averages are $\tsamp$-independent, this is qualitatively different for 
the three ``fluctuations"
\begin{eqnarray}
\muFstar(\tsamp) & \equiv & \beta V \la \overline{\tauhat}^2 \ra \label{eq_muFstar_def} \\
\muF(\tsamp)     & \equiv & \muFtild - \muFstar(\tsamp)          \label{eq_muF_def} \\
\GF(\tsamp)      & \equiv & \muA - \muF(\tsamp) = \GFtild + \muFstar(\tsamp) \label{eq_GF_def} 
\end{eqnarray}
presented in Fig.~\ref{fig_static} which reveal three distinct sampling time regimes.
For very small $\tsamp \ll \tauA \approx 0.13$ (with $\tauA$ being properly defined
below in Sec.~\ref{simu_MSD}) the shear-stress fluctuations $\muF(\tsamp)$ do naturally 
vanish since the instantaneous shear-stress $\tauhat$ has no time to evolve and to 
explore the phase space. (There is no fluctuation for just one data entry.)
Since $\muF(\tsamp) \approx 0$, this implies that $\muFstar(\tsamp) \approx \muFtild$
must have a constant shoulder and the same applies to the generalized stress-fluctuation
formula $\GF(\tsamp)$, Eq.~(\ref{eq_GF_def}). The second time regime of about two orders of magnitude 
between $\tauA$ (left vertical line) and $\tstar \approx 10 \approx 100 \tauA$
(right vertical line) is due to the monotonous decay of $\muFstar(\tsamp)$,
i.e. the ensemble averaged {\em squared} time-averaged shear stress indicated by stars. 
As a consequence, $\muF(\tsamp)$ increases montonously in this interval while $\GF(\tsamp)$ decreases.
The time scale $\tstar$ is the longest (terminal) time scale in this problem.
Due to ergodicity time and ensemble averages become identical in the third sampling
time regime for $\tsamp \gg \tstar$ where $\overline{\tauhat} \to \langle \tauhat \rangle = \tau$, 
$\muFstar(\tsamp) \to \beta V \tau^2$, $\muF(\tsamp) \to \muFtild - \beta V \tau^2$
and $\GF(\tsamp) \to \GF = \Geq$ in agreement with the stress-fluctuation formula, Eq.~(\ref{eq_SFF}). 

\paragraph*{Imposed zero average shear stress.}
Since for convenience we have chosen $\tau = 0$ for our network,
this implies that $\muFstar(\tsamp)$ must vanish and in turn that
$\muF(\tsamp) \to \muFtild$ and $\GF \to \GFtild$ as observed. It is thus
strictly speaking due to the choice $\tau=0$, that the simple
mean $\GFtild$ actually corresponds to the shear modulus $\GF = \Geq$.
For a more general imposed mean shear stress $\tau$ one might,
however, readily use the shifted simple average 
\begin{equation}
\GFtild \Rightarrow \GFtild + \beta V \tau^2 = \GF = \Geq
\label{eq_GFtild_shifted}
\end{equation}
using the known/imposed (not the sampled) $\tau$
as a fast and reliable estimate of the shear modulus $\Geq$
converging several orders of magnitude more rapidly than
the classical (albeit slightly generalized) stress-fluctuation formula $\GF(\tsamp)$. 
We shall come back to the $\tsamp$-dependence of $\muFstar(\tsamp)$
and $\GF(\tsamp)$ in Sec.~\ref{simu_compare}.

\begin{figure}[t]
\centerline{\resizebox{1.0\columnwidth}{!}{\includegraphics*{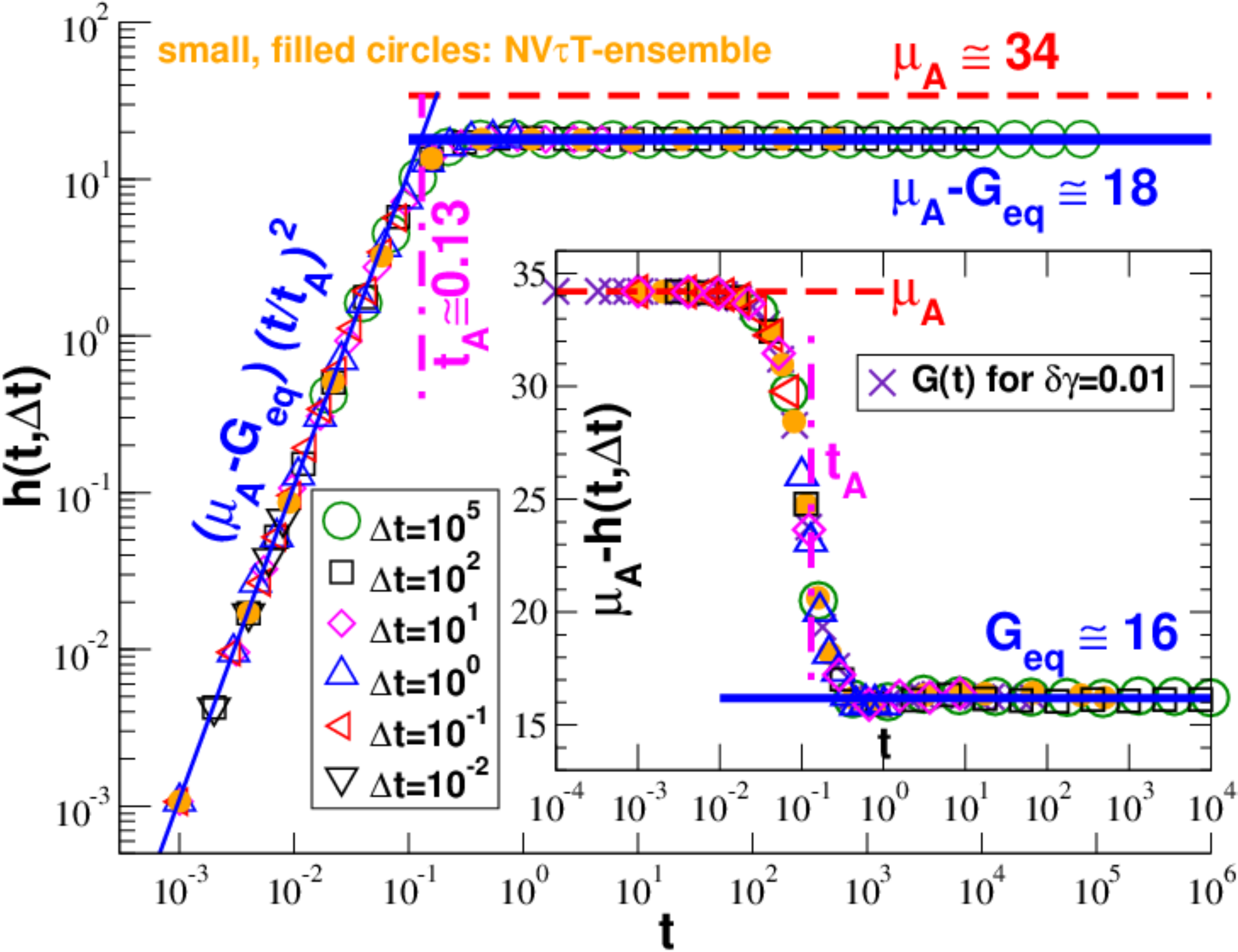}}}
\caption{Shear-stress MSD $h(t,\tsamp)$ as a function of time $t$. 
The filled circles refer to $\NVtT$-ensemble simulations for a sampling time $\tsamp=10^3$,
all open symbols to the corresponding $\NVgT$-ensemble for a broad range of $\tsamp$. 
All data collapse.
Main panel: 
The MSD increases as $h(t) \sim t^2$ for small times $t \ll \tauA$ (thin solid line)
and becomes constant for $t \gg \tauA$ (bold solid line). The crossover time $\tauA \approx 0.13$
is obtained by matching both regimes.
Inset: Comparison of $\muA-h(t)$ with the shear-stress response function $G(t)$ 
obtained from the shear-stress increment $\la \delta \tauhat(t) \ra$ after 
applying a step-strain increment $\delta \gamma = 0.01$ at $t=0$ (crosses). 
\label{fig_MSD}
}
\end{figure}

\subsection{Shear-stress mean-square displacement}
\label{simu_MSD}

\paragraph*{$\tsamp$-independence.}
The MSD $h(t,\tsamp)$ is presented in Fig.~\ref{fig_MSD} as a function of time $t$ 
for a broad range of sampling times $\tsamp$. The data have been computed using 
\begin{equation}
h(t,\tsamp) \equiv \frac{\beta V}{2} 
\la \overline{\left(\tauhat(t+t_0) - \tauhat(t_0) \right)^2} \ra
\mbox{ for } t \le \tsamp
\label{eq_MSD_def}
\end{equation}
where the horizontal bar stands for the gliding average over $t_0$ within a $\tsamp$-package
\cite{AllenTildesleyBook} and $\la \ldots \ra$ for the final ensemble average over the packages.
The first remarkable point in Fig.~\ref{fig_MSD} is the perfect data collapse for all 
sampling times $\tsamp$, i.e. the MSD does not depend explicitly on $\tsamp$. 
This scaling is not surprising since $h(t,\tsamp)$ is a simple average 
measuring the {\em difference} of the shear stresses $\tauhat(t+t_0)$ and $\tauhat(t_0)$
along the trajectory and increasing $\tsamp/\dtMD$ only improves the statistics 
but does not change the expectation value. The second argument $\tsamp$ is dropped from now on.

\paragraph*{Ensemble-independence of MSD.}
The small filled circles in Fig.~\ref{fig_MSD} have been obtained for $\tsamp=10^3$
in the $\NVtT$-ensemble at an imposed average shear stress $\tau = 0$.
An ensemble of $1000$ configurations with quenched shear strains $\gamhat$ 
distributed according to the $\NVtT$-ensemble has been used \cite{WKB15,foot_switchedonbaro}. 
All other data presented have been obtained in the corresponding $\NVgT$-ensemble
\cite{foot_gammazero}.
As already emphasized elsewhere \cite{WKB15}, it is inessential in which ensemble 
we sample the MSD, i.e. 
\begin{equation}
h(t)|_{\tau} = h(t)|_{\gamma} \mbox{ for } t \le \tsamp. 
\label{eq_ht_ensemble}
\end{equation}
The MSD $h(t)$ thus does not transform as a fluctuation,
but as a simple average \cite{AllenTildesleyBook}.
Interestingly, assuming this fundamental scaling property one may (alternatively) demonstrate 
Eq.~(\ref{eq_key}). To see this let us write down the exact identity Eq.~(\ref{eq_htCt}) 
in the $\NVtT$-ensemble 
\begin{equation}
h(t)|_{\tau} = c(0)|_{\tau}-c(t)|_{\tau} = G(0) - G(t) \label{eq_htCt_tau}
\end{equation}
using in the last step that $G(t)=c(t)|_{\tau}$ \cite{foot_ctlimits} 
as shown by integration by parts in Eq.~(15) of Ref.~\cite{WXB15}.
Due to Eq.~(\ref{eq_ht_ensemble}) and $G(0) = \muA = \muF|_{\tau}$ \cite{WXB15},
this directly demonstrates $G(t) = \muA - h(t)|_{\gamma}$ in agreement with Eq.~(\ref{eq_key}).
(For convenience $|_{\gamma}$ is dropped elsewhere.)

\paragraph*{Time-dependence of MSD.}
As seen from the main panel of Fig.~\ref{fig_MSD}, the MSD $h(t)$ of our extremely simple
elastic network shows only two dynamical regimes. For small times $t \ll \tauA \approx 0.13$
the MSD increases quadratically as indicated by the thin solid line. This is to be expected
if the MSD and/or the ACF are analytic around $t=0$ \cite{WKB15,HansenBook}. 
(Strictly speaking, this argument requires time-reversal symmetry, i.e. begs for an 
asymptotically small Langevin thermostat friction constant.)
For larger times $h(t)$ becomes a constant given by $h(t) = \muA-\Geq = \muF$ (bold solid line) in
agreement with Eq.~(\ref{eq_key}) and $G(t) \to \Geq$ for $t \to \infty$.
As seen in the main panel, the crossover time $\tauA$ is determined from the matching of both regimes. 
It marks the time where all forces created by an affine shear transformation
have been relaxed by non-affine displacements \cite{WKB15}.

\paragraph*{Comparison with response function.}
The key relation Eq.~(\ref{eq_key}) is put to the test for all times $t$ in the inset of Fig.~\ref{fig_MSD}.
We compare here $\muA-h(t)$ obtained for different $\tsamp$ with the shear-stress response $G(t)$.
The latter quantity has been computed from the shear-stress increment $\la \delta \tauhat(t) \ra$
with $\delta \tauhat(t) \equiv \tauhat(t) - \tauhat(0^{-})$ measured after a step-strain $\delta \gamma = 0.01$ 
has been applied at $t=0$. This is done using a metric-change of the periodic simulation box 
as illustrated in panel (b) of Fig.~\ref{fig_sketch}.
As in Ref.~\cite{WXB15} $\la \delta \tauhat(t) \ra$ has been averaged over $1000$ configurations. 
The perfect collapse of all data presented confirms the key relation
and this for all sampling times $\tsamp$. This is the main computational result of the present work.

\begin{figure}[t]
\centerline{\resizebox{1.0\columnwidth}{!}{\includegraphics*{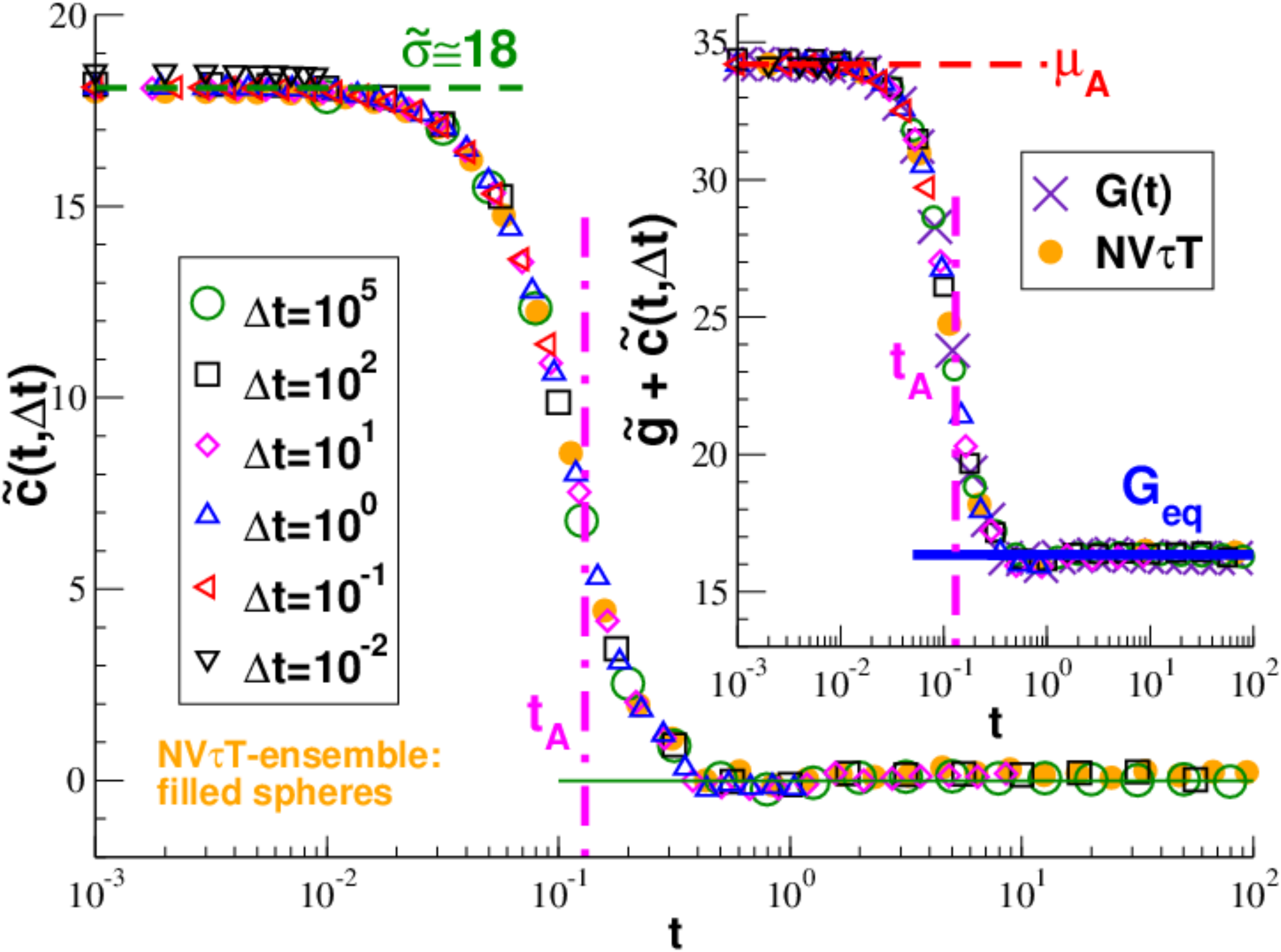}}}
\caption{Shear-stress ACF $\Ctild(t,\tsamp)=\muFtild-h(t,\tsamp)$ as a function 
of time $t$ for different sampling times $\tsamp$ as indicated.
The filled spheres indicate results obtained in the $\NVtT$-ensemble.
As the MSD $h(t)$ the ACF $\Cttild$ is a simple average neither depending on $\tsamp$ or the ensemble.
Main panel: $\Cttild \to \muFtild$ for $t \to 0$ and $\Cttild \to \muFstar = 0$ 
for $t \gg \tauA$.
Inset:
Confirming Eq.~(\ref{eq_GtACFtild}) the shifted ACF $\GFtild + \Cttild$ 
collapses perfectly on the directly measured response function $G(t)$.
\label{fig_ACF}
}
\end{figure}

\subsection{Shear-stress autocorrelation function}
\label{simu_ACF}

We present in Fig.~\ref{fig_ACF} the shear-stress ACF 
\begin{equation}
\Ctild(t,\tsamp) \equiv \beta V 
 \la \overline{\tauhat(t+t_0) \tauhat(t_0)} \ra
\mbox{ for } t \le \tsamp
\label{eq_ACF_def}
\end{equation}
as a function of $t$ comparing different $\tsamp$ for the $\NVgT$-ensemble (open symbols) 
and one example with $\tsamp=10^3$ for the $\NVtT$-ensemble (filled spheres).
Since $\Cttild = \muFtild - h(t)$, this is essentially just a replot of the
data already seen in Fig.~\ref{fig_MSD} using, however, a more common representation.
As one expects the ACF does depend neither on the sampling time nor 
on the ensemble, i.e. $\Cttild$ is a simple average just as $h(t)$.
Interestingly, by writing the identity Eq.~(\ref{eq_htCt}) for the $\NVgT$-ensemble
\begin{equation} 
h(t)|_{\gamma}  = \Ctild(0)|_{\gamma} - \Ctild(t)|_{\gamma} = \muFtild|_{\gamma} - \Ctild(t)|_{\gamma} \label{eq_htCt_gam}
\end{equation}
and using again Eq.~(\ref{eq_ht_ensemble}) and Eq.~(\ref{eq_htCt_tau}) 
one verifies directly that within the $\NVgT$-ensemble 
\begin{eqnarray}
G(t) & = &  \GFtild + \Ctild(t) \label{eq_GtACFtild} \\
     & = &  \GF(\tsamp) + c(t,\tsamp) \label{eq_GtACF}
\end{eqnarray} 
holds where for convenience $|_{\gamma}$ has been dropped 
and the same notations 
$\GF(\tsamp) = \GFtild + \muFstar(\tsamp)$ are used as in Sec.~\ref{simu_static}. 
Since $\GF(\tsamp) \to \GF = \Geq$ for $\tsamp \gg \tstar$, Eq.~(\ref{eq_GtACF}) 
thus confirms Eq.~(\ref{eq_keyPRE})
but generalizes it to finite sampling times $\tsamp$.
As seen in the inset the response function $G(t)$ can be obtained by shifting
$\Cttild \to \GFtild + \Cttild$ confirming thus Eq.~(\ref{eq_GtACF}).
We emphasize that while $\Cttild$ is a sampling-time independent simple average,
the associated ACF $c(t,\tsamp) = \Cttild - \muFstar(\tsamp)$ 
appearing in Eq.~(\ref{eq_htCt_gam}) and Eq.~(\ref{eq_GtACF}) is a {\em fluctuation}. 
It depends on the ensemble \cite{WXB15,WKB15} and on the sampling time 
due to the substracted reference $\muFstar(\tsamp)$.
It is simply for this reason that for $\tsamp \ll \tstar$,
$c(t,\tsamp)$ is numerically less convenient than $\Cttild$.
However, for $\tsamp \gg \tstar$ both ACF become identical (not shown)
since $\Cttild \to \muFstar = \beta V\tau^2 = 0$
due to the zero average shear stress $\tau=0$ chosen.

\subsection{Sampling-time effects revisited}
\label{simu_compare}
We return now to the sampling-time behavior of the fluctuations $\muFstar(\tsamp)$, 
$\muF(\tsamp)$ and $\GF(\tsamp)$ shown in Fig.~\ref{fig_static}. 
As seen, e.g., from Eq.~(20) of Ref.~\cite{WXB15}, the $\tsamp$-effects can be 
understood by noticing that $\muF(\tsamp)$ may be written as 
a weighted integral over the MSD $h(t)$ \cite{foot_polymer}
\begin{equation}
\muF(\tsamp) = \frac{2}{\tsamp^2} \int_0^{\tsamp} \ \ddiff t \ (\tsamp-t) h(t).
\label{eq_muFtsamp_ht}
\end{equation}
Time translational invariance is assumed here and we have used that the MSD does not
explicitly depend $\tsamp$. The sampling-time dependence of $\muF(\tsamp)$ is thus
reduced to the time dependence of $h(t)$. Equation~(\ref{eq_muFtsamp_ht})
and the corresponding relations for $\GF(\tsamp) = \muA - \muF(\tsamp)$ and 
$\muFstar(\tsamp) = \muFtild - \muF(\tsamp)$ are indicated by thin dashed lines 
in Fig.~\ref{fig_static}.
Also given is the simple-average expression $\muA - h(t=\tsamp)$ for $\tsamp \le 10^2$ (diamonds)
using only the end-points of $\tsamp$-packages which are then in addition ensemble-averaged 
over $10^4$ independent packages. Since this corresponds to the response modulus $G(t)$ taken
at $t=\tsamp$, it converges to $\Geq$ already for sampling times $\tsamp \gg \tauA$,
i.e. two orders of magnitude earlier than the stress-fluctuation formula $\GF(\tsamp)$
which only converges for $\tsamp \gg \tstar \approx 100 \tauA$.
The reason for this stems simply from the inequality
\begin{equation}
h(t = \tsamp) \ge \frac{2}{\tsamp^2} \int_0^{\tsamp} \ \ddiff t \ (\tsamp-t) h(t) 
\label{eq_inequality}
\end{equation}
for a monotonously increasing function $h(t)$. Unfortunately, assuming
the same number of $\tsamp$-packages, $\muA - h(\tsamp)$ fluctuates
much more strongly than $\GF(\tsamp)$, just as the end-to-end distance
of a polymer chain fluctuates much more strongly as its radius of gyration.
Equation~(\ref{eq_key}) is thus only of interest for the determination of $\Geq$
if a large ensemble of short trajectories with $\tauA \ll \tsamp \ll \tstar$
has been computed. As already pointed out in Sec.~\ref{simu_static},
the most efficient property for the computation of the modulus is
the simple average $\GFtild + \beta V \tau^2$.
We note finally that since $h(t)$ is characterized by only one characteristic time,
the crossover time $\tauA$, Eq.~(\ref{eq_muFtsamp_ht}) implies that the terminal time $\tstar$ 
must be a unique function of $\tauA$ \cite{foot_workedout}. 
Our simple network is thus only characterized by {\em one} characteristic time. 

\section{Conclusion}
\label{sec_conc}

\paragraph*{Summary.}
Rewriting the central relation Eq.~(\ref{eq_keyPRE}) of Ref.~\cite{WXB15}
it is shown that the shear-relaxation modulus $G(t)$ of an isotropic elastic 
body may be computed as $G(t) = \muA - h(t)$ 
in terms of the difference of the two simple averages $\muA$ and $h(t)$ characterizing,
respectively, the canonical-affine strain response $G(0)$ at $t=0$ and the subsequent 
stress relaxation process for $t > 0$.
Interestingly, Eq.~(\ref{eq_keyPRE}) and Eq.~(\ref{eq_key}) may be directly demonstrated from 
the fundamental scaling $h(t)|_{\gamma}=h(t)|_{\tau}$ (Sec.~\ref{simu_MSD}).
Note that $G(t) = \muA - h(t)$ and $G(t) = \GFtild + \Ctild(t)$ are equivalent 
simple-average expressions as shown, respectively in Fig.~\ref{fig_MSD} and Fig.~\ref{fig_ACF}. 
From the practical point of view it is important that $h(t)$ or $\Cttild$ do not depend explicitly 
on the sampling time $\tsamp$ and the response function $G(t)$ may thus be computed even if $\tsamp$ 
is much smaller than the terminal relaxation time $\tstar$ of the system. 
(For our simple networks $\tstar \approx 100 \tauA$ with $\tauA$ being the crossover time of the MSD.)
As shown in Sec.~\ref{simu_ACF}, the relaxation modulus may be also obtained from the ACF 
$c(t,\tsamp) = \Cttild - \muFstar(\tsamp)$ using $G(t) = \GF(\tsamp) + c(t,\tsamp)$
with $\GF(\tsamp) = \GFtild + \muFstar(\tsamp)$ being the generalized, sampling-time 
dependent stress-fluctuation estimate for the shear modulus $\Geq$. 
For $\tsamp \gg \tstar$ these relations reduce to Eq.~(\ref{eq_keyPRE}).
Finally, comparing $\muA - h(t \approx \tsamp)$ with the stress-fluctuation 
formula $\GF(\tsamp)$ it has been shown (Sec.~\ref{simu_compare}) that the former expression 
converges about two orders of magnitude more rapid, albeit with lesser accuracy
depending on the number of independent $\tsamp$-packages used.
The fastest convergence has been obtained, however, using the simple average 
$\GFtild + \beta V \tau^2$ (Fig.~\ref{fig_static}). 

\paragraph*{Discussion.}
While the present paper has focused on solids, it should be emphasized that Eq.~(\ref{eq_key}), 
being derived using quite general arguments not relying on a well-defined particle displacement field, 
should apply also to systems with plastic rearrangements and to the liquid limit.
Due to its $\tsamp$-independence it should be useful especially 
for complex liquids and glass-forming systems \cite{WittenPincusBook}
with computationally non-accessible terminal relaxation times $\tstar$.
We emphasize that the commonly used expression $G(t) = \Ctild(t)$ requires $\GFtild=\muA-\muFtild=0$ 
to hold \cite{foot_liquid}. While this condition is justified for a liquid where 
$\GFtild = \GF = \Geq = 0$,
it is {\em incorrect} in general as shown by the example presented in this work 
(Fig.~\ref{fig_static}). 
Hence, some care is needed when approximating $G(t)$ by $\Cttild$ for systems below the glass transition
\cite{Klix12,Szamel15}. 
Since Eq.~(\ref{eq_key}) can be used in any case and since it is not much more difficult to compute,
it provides a rigorous alternative without additional assumptions \cite{Allen94,foot_eta}.

\paragraph*{Outlook.}
Naturally, one expects that Eq.~(\ref{eq_key}) can be generalized for more general linear 
relaxation moduli $M(t)$ of classical elastic bodies and fluids. With 
\begin{equation}
\MA = V \la \left. \partial \hat{i}_{\beta}/\partial x_{\alpha} \right|_{x_{\alpha}x_{\beta}} \ra
\label{eq_MA}
\end{equation}
characterizing the initial canonical-affine response $M(0)$ of the system to an infinitesimal change 
$\delta x_{\alpha}$ of an extensive variable $x_{\alpha}$ and $\delta i_{\beta}$ the subsequent change 
of an intensive system variable $i_{\beta}$ and 
\begin{equation}
h(t) \equiv \frac{\beta V}{2} \la \overline{\left(\hat{i}_{\alpha}(t)-\hat{i}_{\alpha}(0)\right) 
                                  \left(\hat{i}_{\beta}(t)-\hat{i}_{\beta}(0)\right)} \ra
\label{eq_MSDgen}
\end{equation}
being the generalized MSD associated with the instantaneous intensive variables 
$\hat{i}_{\alpha}(t)$ and $\hat{i}_{\beta}(t)$, one expects $h(t)$ to be a simple average,
i.e. $h(t)|_{x_{\alpha}x_{\beta}} = h(t)|_{i_{\alpha}i_{\beta}}$, and a generalized simple-average expression
\begin{equation}
M(t) = M(0) - h(t) \mbox{ with } M(0) = \MA
\label{eq_key_gen}
\end{equation} 
should thus hold again.
The reformulation of the general stress-fluctuation formalism in terms of such simple averages 
and the test of its computational efficiency are currently under way.

\vspace*{0.2cm} 
\begin{acknowledgments}
H.X. thanks the IRTG Soft Matter for financial support.
We are indebted to H. Meyer (Strasbourg) and A.E. Likhtman (Reading) 
for helpful discussions.
\end{acknowledgments}


\begin{thebibliography}{23}
\expandafter\ifx\csname natexlab\endcsname\relax\def\natexlab#1{#1}\fi
\expandafter\ifx\csname bibnamefont\endcsname\relax
  \def\bibnamefont#1{#1}\fi
\expandafter\ifx\csname bibfnamefont\endcsname\relax
  \def\bibfnamefont#1{#1}\fi
\expandafter\ifx\csname citenamefont\endcsname\relax
  \def\citenamefont#1{#1}\fi
\expandafter\ifx\csname url\endcsname\relax
  \def\url#1{\texttt{#1}}\fi
\expandafter\ifx\csname urlprefix\endcsname\relax\def\urlprefix{URL }\fi
\providecommand{\bibinfo}[2]{#2}
\providecommand{\eprint}[2][]{\url{#2}}

\bibitem[{\citenamefont{Rubinstein and Colby}(2003)}]{RubinsteinBook}
\bibinfo{author}{\bibfnamefont{M.}~\bibnamefont{Rubinstein}} \bibnamefont{and}
  \bibinfo{author}{\bibfnamefont{R.}~\bibnamefont{Colby}},
  \emph{\bibinfo{title}{Polymer Physics}} (\bibinfo{publisher}{Oxford
  University Press}, \bibinfo{address}{Oxford}, \bibinfo{year}{2003}).

\bibitem[{\citenamefont{Witten and Pincus}(2004)}]{WittenPincusBook}
\bibinfo{author}{\bibfnamefont{T.}~\bibnamefont{Witten}} \bibnamefont{and}
  \bibinfo{author}{\bibfnamefont{P.~A.} \bibnamefont{Pincus}},
  \emph{\bibinfo{title}{Structured Fluids: Polymers, Colloids, Surfactants}}
  (\bibinfo{publisher}{Oxford University Press}, \bibinfo{address}{Oxford},
  \bibinfo{year}{2004}).

\bibitem[{\citenamefont{Doi and Edwards}(1986)}]{DoiEdwardsBook}
\bibinfo{author}{\bibfnamefont{M.}~\bibnamefont{Doi}} \bibnamefont{and}
  \bibinfo{author}{\bibfnamefont{S.~F.} \bibnamefont{Edwards}},
  \emph{\bibinfo{title}{The Theory of Polymer Dynamics}}
  (\bibinfo{publisher}{Clarendon Press}, \bibinfo{address}{Oxford},
  \bibinfo{year}{1986}).

\bibitem[{\citenamefont{Allen and Tildesley}(1994)}]{AllenTildesleyBook}
\bibinfo{author}{\bibfnamefont{M.}~\bibnamefont{Allen}} \bibnamefont{and}
  \bibinfo{author}{\bibfnamefont{D.}~\bibnamefont{Tildesley}},
  \emph{\bibinfo{title}{Computer Simulation of Liquids}}
  (\bibinfo{publisher}{Oxford University Press}, \bibinfo{address}{Oxford},
  \bibinfo{year}{1994}).

\bibitem[{\citenamefont{Squire et~al.}(1969)\citenamefont{Squire, Holt, and
  Hoover}}]{Hoover69}
\bibinfo{author}{\bibfnamefont{D.~R.} \bibnamefont{Squire}},
  \bibinfo{author}{\bibfnamefont{A.~C.} \bibnamefont{Holt}}, \bibnamefont{and}
  \bibinfo{author}{\bibfnamefont{W.~G.} \bibnamefont{Hoover}},
  \bibinfo{journal}{Physica} \textbf{\bibinfo{volume}{42}},
  \bibinfo{pages}{388} (\bibinfo{year}{1969}).

\bibitem[{\citenamefont{Barrat et~al.}(1988)\citenamefont{Barrat, Roux, Hansen,
  and Klein}}]{Barrat88}
\bibinfo{author}{\bibfnamefont{J.-L.} \bibnamefont{Barrat}},
  \bibinfo{author}{\bibfnamefont{J.-N.} \bibnamefont{Roux}},
  \bibinfo{author}{\bibfnamefont{J.-P.} \bibnamefont{Hansen}},
  \bibnamefont{and} \bibinfo{author}{\bibfnamefont{M.~L.} \bibnamefont{Klein}},
  \bibinfo{journal}{Europhys. Lett.} \textbf{\bibinfo{volume}{7}},
  \bibinfo{pages}{707} (\bibinfo{year}{1988}).

\bibitem[{\citenamefont{Barrat}(2006)}]{Barrat06}
\bibinfo{author}{\bibfnamefont{J.-L.} \bibnamefont{Barrat}}, in
  \emph{\bibinfo{booktitle}{Computer Simulations in Condensed Matter Systems:
  From Materials to Chemical Biology}}, edited by
  \bibinfo{editor}{\bibfnamefont{M.}~\bibnamefont{Ferrario}},
  \bibinfo{editor}{\bibfnamefont{G.}~\bibnamefont{Ciccotti}}, \bibnamefont{and}
  \bibinfo{editor}{\bibfnamefont{K.}~\bibnamefont{Binder}}
  (\bibinfo{publisher}{Springer}, \bibinfo{address}{Berlin and Heidelberg},
  \bibinfo{year}{2006}), vol. \bibinfo{volume}{704}, pp.
  \bibinfo{pages}{287---307}.

\bibitem[{\citenamefont{Lutsko}(1989)}]{Lutsko89}
\bibinfo{author}{\bibfnamefont{J.~F.} \bibnamefont{Lutsko}},
  \bibinfo{journal}{J. Appl. Phys} \textbf{\bibinfo{volume}{65}},
  \bibinfo{pages}{2991} (\bibinfo{year}{1989}).

\bibitem[{\citenamefont{Flenner and Szamel}(2015)}]{Szamel15}
\bibinfo{author}{\bibfnamefont{E.}~\bibnamefont{Flenner}} \bibnamefont{and}
  \bibinfo{author}{\bibfnamefont{G.}~\bibnamefont{Szamel}},
  \bibinfo{journal}{Phys. Rev. Lett.} \textbf{\bibinfo{volume}{107}},
  \bibinfo{pages}{105505} (\bibinfo{year}{2015}).

\bibitem[{\citenamefont{Wittmer et~al.}(2013)\citenamefont{Wittmer, Xu,
  Poli\'nska, Weysser, and Baschnagel}}]{WXP13}
\bibinfo{author}{\bibfnamefont{J.~P.} \bibnamefont{Wittmer}},
  \bibinfo{author}{\bibfnamefont{H.}~\bibnamefont{Xu}},
  \bibinfo{author}{\bibfnamefont{P.}~\bibnamefont{Poli\'nska}},
  \bibinfo{author}{\bibfnamefont{F.}~\bibnamefont{Weysser}}, \bibnamefont{and}
  \bibinfo{author}{\bibfnamefont{J.}~\bibnamefont{Baschnagel}},
  \bibinfo{journal}{J. Chem. Phys.} \textbf{\bibinfo{volume}{138}},
  \bibinfo{pages}{12A533} (\bibinfo{year}{2013}).

\bibitem[{\citenamefont{Wittmer
  et~al.}(2015{\natexlab{a}})\citenamefont{Wittmer, Xu, and
  Baschnagel}}]{WXB15}
\bibinfo{author}{\bibfnamefont{J.~P.} \bibnamefont{Wittmer}},
  \bibinfo{author}{\bibfnamefont{H.}~\bibnamefont{Xu}}, \bibnamefont{and}
  \bibinfo{author}{\bibfnamefont{J.}~\bibnamefont{Baschnagel}},
  \bibinfo{journal}{Phys. Rev. E} \textbf{\bibinfo{volume}{91}},
  \bibinfo{pages}{022107} (\bibinfo{year}{2015}{\natexlab{a}}).

\bibitem[{\citenamefont{Wittmer
  et~al.}(2015{\natexlab{b}})\citenamefont{Wittmer, Kriuchevskyi, Baschnagel,
  and Xu}}]{WKB15}
\bibinfo{author}{\bibfnamefont{J.~P.} \bibnamefont{Wittmer}},
  \bibinfo{author}{\bibfnamefont{I.}~\bibnamefont{Kriuchevskyi}},
  \bibinfo{author}{\bibfnamefont{J.}~\bibnamefont{Baschnagel}},
  \bibnamefont{and} \bibinfo{author}{\bibfnamefont{H.}~\bibnamefont{Xu}},
  \bibinfo{journal}{Eur. Phys. J. B} \textbf{\bibinfo{volume}{88}},
  \bibinfo{pages}{242} (\bibinfo{year}{2015}{\natexlab{b}}).

\bibitem[{\citenamefont{Lebowitz et~al.}(1967)\citenamefont{Lebowitz, Percus,
  and Verlet}}]{Lebowitz67}
\bibinfo{author}{\bibfnamefont{J.~L.} \bibnamefont{Lebowitz}},
  \bibinfo{author}{\bibfnamefont{J.~K.} \bibnamefont{Percus}},
  \bibnamefont{and} \bibinfo{author}{\bibfnamefont{L.}~\bibnamefont{Verlet}},
  \bibinfo{journal}{Phys. Rev.} \textbf{\bibinfo{volume}{153}},
  \bibinfo{pages}{250} (\bibinfo{year}{1967}).

\bibitem[{\citenamefont{Hansen and McDonald}(2006)}]{HansenBook}
\bibinfo{author}{\bibfnamefont{J.}~\bibnamefont{Hansen}} \bibnamefont{and}
  \bibinfo{author}{\bibfnamefont{I.}~\bibnamefont{McDonald}},
  \emph{\bibinfo{title}{Theory of simple liquids}}
  (\bibinfo{publisher}{Academic Press}, \bibinfo{address}{New York},
  \bibinfo{year}{2006}), \bibinfo{note}{3nd edition}.

\bibitem[{\citenamefont{Klix et~al.}(2012)\citenamefont{Klix, Ebert, Weysser,
  Fuchs, Maret, and Keim}}]{Klix12}
\bibinfo{author}{\bibfnamefont{C.}~\bibnamefont{Klix}},
  \bibinfo{author}{\bibfnamefont{F.}~\bibnamefont{Ebert}},
  \bibinfo{author}{\bibfnamefont{F.}~\bibnamefont{Weysser}},
  \bibinfo{author}{\bibfnamefont{M.}~\bibnamefont{Fuchs}},
  \bibinfo{author}{\bibfnamefont{G.}~\bibnamefont{Maret}}, \bibnamefont{and}
  \bibinfo{author}{\bibfnamefont{P.}~\bibnamefont{Keim}},
  \bibinfo{journal}{Phys. Rev. Lett.} \textbf{\bibinfo{volume}{109}},
  \bibinfo{pages}{178301} (\bibinfo{year}{2012}).

\bibitem[{foo({\natexlab{a}})}]{foot_gammazero}
\bibinfo{note}{The specific network we use has a small, but non-vanishing shear
  stress $\tau \approx 0.012$ at a fixed strain $\gamma=0$. If the strain is
  allowed to fluctuate freely in the $\NVtT$-ensemble at an imposed average
  shear stress $\tau = 0$ this yields a tiny, non-vanishing average shear
  strain $\gamma_0=0.00071$ \cite{WKB15}. Since we wish to compare stress
  fluctuations in the $\NVtT$- and the $\NVgT$-ensemble at the same state point
  with $\tau=0$, all $\NVgT$-ensemble simulations are performed at $\gamma_0$
  as in Ref.~\cite{WKB15}. The differences with respect to simulations at
  $\gamma=0$ are obviously negligible.}

\bibitem[{foo({\natexlab{b}})}]{foot_switchedonbaro}
\bibinfo{note}{As described in more detail in Ref.~\cite{WKB15}, a similar
  result is obtained using a very slow switched-on shear-barostat. If a strong
  shear-barostat is applied, $h(t)$ approaches instead rapidly $\muFtau =
  \muA$.}

\bibitem[{foo({\natexlab{c}})}]{foot_ctlimits}
\bibinfo{note}{The limit $\tsamp \gg \tstar$ has been taken for $c(t,\tsamp)$
  in Eq.~(\ref{eq_htCt_tau}). This is allowed since $h(t)$ does not depend on
  $\tsamp$. The limit is required by the thermodynamic argument leading to
  $G(t)=c(t)|_{\tau}$.}

\bibitem[{foo({\natexlab{d}})}]{foot_polymer}
\bibinfo{note}{A relation similar to Eq.~(\ref{eq_muFtsamp_ht}) exists in
  polymer theory \cite{DoiEdwardsBook} expressing the radius of gyration of a
  polymer chain as a weighted integral over internal mean-squared segment sizes
  \cite{DoiEdwardsBook}.}

\bibitem[{foo({\natexlab{e}})}]{foot_workedout}
\bibinfo{note}{The ratio $\tstar/\tauA \approx 100$ may be worked out by
  analyzing the integral $\int_0^x \ddiff s (x-s) h(s) 2/x^2$ with
  $x=\tsamp/\tauA$ and $s=t/\tsamp$ and using that $h(s)=\muF s^2$ for $s < 1$
  and $h(s) = \muF$ for $s > 1$.}

\bibitem[{foo({\natexlab{f}})}]{foot_liquid}
\bibinfo{note}{According to Eq.~(\ref{eq_GF_def}) this condition implies
  $\GF(\tsamp) = \muFstar(\tsamp)$ for the generalized stress-fluctuation
  formula for a sampling-time dependent shear modulus.}

\bibitem[{\citenamefont{Allen et~al.}(1994)\citenamefont{Allen, Brown, and
  Masters}}]{Allen94}
\bibinfo{author}{\bibfnamefont{M.}~\bibnamefont{Allen}},
  \bibinfo{author}{\bibfnamefont{D.}~\bibnamefont{Brown}}, \bibnamefont{and}
  \bibinfo{author}{\bibfnamefont{A.}~\bibnamefont{Masters}},
  \bibinfo{journal}{Phys. Rev. E} \textbf{\bibinfo{volume}{49}},
  \bibinfo{pages}{2488} (\bibinfo{year}{1994}).

\bibitem[{foo({\natexlab{g}})}]{foot_eta}
\bibinfo{note}{The linear shear viscosity $\eta$ may be obtained by integrating
  over $G(t)=\muA-h(t)$. Note that this integral has a different form as the
  Einstein relations for the shear-stress response discussed in the literature
  \cite{AllenTildesleyBook,Allen94} and that it does {\em not} suffer from
  problems related to the periodic boundary conditions.}

\end{thebibliography}

\end{document}